\documentclass[%
 reprint,
nofootinbib,
 amsmath,amssymb,
 aps,
pra,
]{revtex4-2}
\usepackage{longtable}
\usepackage{booktabs}

\usepackage{graphicx}% Include figure files
\usepackage{dcolumn}% Align table columns on decimal point
\usepackage{bm}% bold math
\usepackage{braket}
\begin{document}
\newcommand{\thistitle}{The building blocks of asymptotically flat spacetimes}
\title{\thistitle}
\author{Shadi Ali Ahmad}
\affiliation{
Center for Cosmology and Particle Physics,\\
New York University,\\
New York, NY 10003, USA}
\date{\today}
\begin{abstract}
The classification of all possible induced representations arising from theories admitting a Poincaré symmetry has molded our very conception of particles in flat space. In this short note, we argue that if one takes the same viewpoint on the \textit{asymptotic} symmetry group instead of the global one, the familiar relativistic classification is enhanced to provide a new notion of \textit{excitations} in asymptotically flat spacetimes. These excitations blur the line between the quantum description of a particle and a spacetime geometry, and will appear in \textit{any} theory with the same symmetry. By interacting together, they produce both the geometry \textit{of} and familiar propagating particles \textit{in} spacetime.
\end{abstract}
\maketitle
\section{Motivation} In the search for a fundamental description of physical phenomena, an organizing principle is essential. Symmetry laws can serve as this principle. The classification of all basic or irreducible representations of a given symmetry allows us to reconstruct the specifics of \textit{any} theory admitting that symmetry. Once the building blocks are known, one simply has to organize them appropriately to describe a specific theory. 

Perhaps the most familiar example of the power of this perspective is Wigner's classification of relativistic particles by studying the representations of the Poincaré group $P$~\cite{wigner_gruppentheorie_1931, Wigner:1939cj}. Not only do we obtain fermions and bosons, but we also learn about other systems. For example, the electrons of graphene near certain critical points enjoy an emergent relativistic symmetry~\cite{graphene}. The fact that representation theory has something to say about any such theory with the same symmetry is a general phenomenon. 

In this note, we will be interested in a symmetry much like the Poincaré group, but infinitely richer. To set the stage, we consider groups of the form 
\begin{equation}
     G_{\mathcal{S}} :=\text{Diff}(\mathcal{S}) \rtimes A^{\mathcal{S}}, \nonumber
\end{equation}
where the first is the diffeomorphism group of a surface $\mathcal{S}$ acting on the second factor, and the second is comprised of sections $\alpha : \mathcal{S} \to A$ with $A$ being an Abelian group under addition. This group and its related family members have surfaced recently in a variety of physical systems as diverse as hydrodynamics~\cite{baguis1998semidirect, holm1998euler,holm2002euler, khesin2020geometric, donnelly2021gravitational}~and asymptotically flat spacetimes~\cite{bondi1962gravitational,sachs1962gravitational,sachs1962asymptotic}, including the corner proposal of gravitational subregion physics ~\cite{donnelly2016local, speranza2018local, geiller2018lorentz, freidel2020edge1,freidel2020edge2,freidel2021edge3,donnelly2021gravitational, ciambelli2021isolated, freidel2021extended, ciambelli2022embeddings}, and anyonic parastatistics~\cite{leinaas1977theory, goldin1980particle, goldin1981representations, wilczek1982magnetic}. In an effort to correct the scientific record on the subject of anyons, we refer the reader to a historical review of the anyon, written by one of its discoverers~\cite{Goldin:2022gcn}.

In the context of gravity, taking the surface to be the sphere and $A$ to be the translation group $\mathbb{R}$, identifies the above group with an infinite-dimensional extension of the Poincaré group. We refer to this group, denoted $B$, as the \textit{asymptotic} symmetry group of flat space. Its elements are superrotations\footnote{A more apt name for transformations that extend the Lorentz subgroup $L$ would be supra-Lorentz transformations, which avoids implying that they only generalize rotations \textit{and} distinguishes them from supersymmetric extensions.} $\Lambda \in \text{Diff}(S^2)$ and supertranslations $\alpha \in C^{\infty}(S^2)$. The dual of a supertranslation will be called a supermomentum $\mathcal{P}(x)$, where $x \in S^2$. It is important to note that this classification has been performed for $B_{0} := L \rtimes C^{\infty}(S^2)$, where $L \simeq SO(3,1)$ is the Lorentz group~\cite{mccarthy0,mccarthy1,mccarthy2,mccarthy3}.\footnote{Since projective representations are those relevant for quantum theory, it is more appropriate to consider the unitary representations of the universal cover of a given group. For the Lorentz group $L$, this is $\tilde{L} := SL_{2}(\mathbb{C})$.} The mathematical results were interpreted physically quite recently~\cite{bekaert2024bmsparticles,Bekaert:2025kjb}, with new insights related to the degeneracy of vacua and the memory effect by way of the infrared triangle~\cite{Strominger:2017zoo}.

Before we begin, we comment on two possible sources of confusion. First, the diffeomorphism group is comprised of maps $\Lambda : S^2 \to S^2$, which may or may not be \textit{isometries}. Isometries are diffeomorphisms which \textit{do} preserve the metric on the sphere. Second, we refer to the representations as \textit{excitations}, since in paraphrasing Haag~\cite{haag1996local}, particles are corollaries and not axioms. We will only reserve the term particle for a \textit{unitary irreducible representation of the Poincaré group}. More generally, these could emerge from fields or a higher structure \textit{defining} the theory like the operator product expansion~\cite{Wilson1969, Zimmermann1973, HollandsWald2001}.

 The purpose of this note is to report on a new notion of excitations, which we term \textit{asymptotic}, by studying the induced representations of the asymptotic symmetry group $B$. We will argue that they are the building blocks of asymptotically flat spacetimes. Along the way, we will find that we recover the familiar particles without tachyons, but they will have larger configuration spaces than even the BMS `particles' obtained in Refs.~\cite{mccarthy0,mccarthy1,mccarthy2,mccarthy3, bekaert2024bmsparticles, Bekaert:2025kjb}. It is important to mention that previous investigations, in four dimensions, did not deal with superrotations. One can summarize our result succinctly: just as reckoning with supertranslations revealed novel irreducibles which restrict to Poincaré reducibles~\cite{mccarthy2}, accounting for superrotations forces those same irreducibles to become reducible upon turning off the superrotations. The usual particles become reducible sums of the building blocks, but these asymptotic excitations have more to say about spacetime than the Poincaré particles. Indeed, it will be shown how these novel asymptotic excitations distort the distinction between geometry and matter, propelling us, out of necessity, towards quantum gravity. The reasons these representations were overlooked are given below and a survey of exciting potential prospects are described.

\section{Induced particles or excitations?}
We review the philosophy of the inducing construction, due originally to Mackey~\cite{Mackey1970, mackey1976theory} and refined over the years with further applications to physics~\cite{rawnsley1975representations, marsden1984reduction, marsden1984semidirect,knapp2013lie}. It will hopefully be clear that this construction is simple yet very powerful. As an \textit{application}, it is able to recover particles detected in the physical world.

We will work in the context of a semidirect product group $G \rtimes A$, where $G$ is a Lie group acting on the Abelian group $A$. To faciliate the presentation of our results, we will generically refer to elements of $G$ as superrotations and elements of $A^{*}$, the dual of $A$, as supermomenta. As a byproduct, we will also classify all the \textit{dressed} excitations of the asymptotic symmetry group $B$ which can be interpreted in terms of relativistic particles with additional degrees of freedom. 

The refrain goes: begin with a character, or physically a supermomentum $\mathcal{P}(x) \in A^{*}$, the dual space of supertranslations. Supermomenta pair with supertranslations under the inner product $\braket{\mathcal{P}| \alpha} := \int_{S^{2}} d\Omega \mathcal{P}(x) \alpha(x)$. Here, $d\Omega = \sin \theta d \theta d \phi$ is the volume form of the round sphere. This allows us to obtain the action of superrotations on supermomenta, by duality $\braket{\mathcal{P}|g \cdot \alpha} = \braket{g^{-1}\cdot  \mathcal{P}|\alpha}$. This defines the dual action of $G$ on $\mathcal{P}$. The orbit obtained can then be described as the homogenous space for $G$ with stabilizer $G_{\mathcal{P}}  \leqslant G$. The stabilizer, which is physically referred to as the little group, is the subgroup of $G$ that preserves the supermomentum, $g \cdot \mathcal{P} = \mathcal{P}$.

To classify the excitations of a theory charged under $B$, we thus need to obtain all induced\footnote{Under certain regularity conditions, it can be shown that all such excitations are induced~\cite{BarutRaczka1986}.} representations starting from $A$. These are differentiated by orbits of supermomenta $O_{\mathcal{P}}$ and spin spaces $\mathfrak{H}$ on which the stabilizer $G_{\mathcal{P}}$ is represented. The configuration space of a given representation is 
\begin{equation}
 O_{\mathcal{P}} \otimes \mathfrak{H} \simeq G / G_{\mathcal{P}} \otimes \mathfrak{H}, \nonumber
\end{equation}
on top of which a Hilbert space $L^{2}_{\mathcal{P}}(\mu)$ can be constructed as square-integrable wavefunctions valued in the configuration space above, with a quasi-invariant measure $\mu$ associated to the orbit. 

To contextualize our result, we point out some potential shortcomings of restricting too soon to a (non-unique) Poincaré subgroup $P$ of $B$. For that purpose, we assume in this section that the supermomentum $\mathcal{P}$ decomposes into a hard four-momentum $p$ and a soft piece $b$, in line with the dominant decomposition in the literature
\begin{equation}
    \mathcal{P}(x) =  p^{\mu} + b(\theta, \phi), \nonumber
\end{equation}
where we have highlighted the generic dependence on both coordinates on the sphere. 

Any transformation fixing $\mathcal{P}$ must also fix $p$, in other words, $G_{\mathcal{P}} \leqslant G_{p}$. If one takes $G \simeq L$, this effectively turns off the superrotation field. This is precisely the setting of Refs.~\cite{mccarthy0,mccarthy1,mccarthy2,mccarthy3} and Ref.~\cite{bekaert2024bmsparticles, Bekaert:2025kjb}. However, even when $G = \text{Diff}(S^2)$, the assumption of non-trivial hard and soft parts of $\mathcal{P}$ forces the same containment of the stabilizers $G_{\mathcal{P}} \leqslant G_{p} \leqslant \text{Diff}(S^2)$. 

After making this assumption clear, one can firmly distinguish the induced representations from the standard Poincaré particles. The configuration space associated to a given orbit $O_{\mathcal{P}}$ may be related to that of $p$. The obtained excitations then have the following schematic orbit
\begin{equation}
        O_{\mathcal{P}} \simeq \text{Diff}(S^2)/ G_{\mathcal{P}} \simeq \left[ \text{Diff}(S^2)/ L \times  G_{p} / G_{\mathcal{P}} \right] \times  L / G_{p} , \nonumber
    \end{equation}
    where $\text{Diff}(S^2) / L$ encodes an infinite-dimensional degeneracy in superrotation vacua, $G_{p} / G_{\mathcal{P}}$ respects the misalignment of the stabilizers and encodes supertranslation vacua, and $O_{p} := L / G_{p}$ is identified with the orbit of the hard part $p$ under Lorentz transformations. The space in brackets thus serves to quantify the enhancement from the Poincaré group $P$ to the asymptotic symmetry group $B$. One can view these labels as those of a fixed spacetime, with the specific Poincaré particle $p$ propagating in inequivalent geometries with non-trivial superrotations and supertranslations turned on. Generically, this particle $O_{p}$ is reducible relative to $B$. The branching of the spin labels arising from $G_{\mathcal{P}}$ into those of $G_{p}$ may be explicitly worked out. This reorients the references mentioned previously within our context.

\section{Asymptotic excitations}
    The infinite-dimensional label arising from $\text{Diff}(S^2) / L$ should be alarming, as these are typically thought of as irreducible representations. The orbit is too large, and this label is independent of $\mathcal{P}$. Motivated by this, we use a different decomposition\footnote{It is important to note that both quoted decompositions generate the entire space of supermomenta $C^{\infty}(S^2)$.} of the supermomentum 
    \begin{equation}
        \mathcal{P}(\theta, \phi) = \rho + j(\theta) + s(\theta, \phi), \nonumber
    \end{equation}
    where the first term is a constant density, the second is an axisymmetric function of the sphere, and the last is a mixed function of both angles. The decomposition above striates the supermomenta according to particular symmetries they may possess, relative to $G \simeq \text{Diff}(S^2)$. In analogy to obtaining irreducibles of Poincaré or $B_{0}$, the strategy is to classify the stabilizers and orbits of each of the above classes of supermomenta. To do that, we first recall that the diffeomorphism group of the sphere is \textit{diffeomorphic}\footnote{In fact, this holds for any compact oriented surface. This will be relevant for topology change. Moreover, this decomposition does not generically respect the group structure.} to
\begin{equation}
    \text{Diff}(S^2) \simeq \text{SDiff}(S^2) \times \mathcal{W}, \nonumber
\end{equation}
where the first factor is comprised of \textit{special} diffeomorphisms with unit Jacobian and the second is comprised of Weyl factors $\omega (\theta, \phi)$ on the sphere~\cite{omori_group_1970,Ebin1970GroupsOD,La:1995mc}. This decomposition relies on a choice of some canonical volume form on $S^{2}$, $d\Omega$, and any Weyl factor $\omega$ must give rise to the same total volume as the canonical one.

To be more explicit about the action on supermomenta, which is crucial for this analysis, we have that
\begin{equation}
   \Lambda \cdot \mathcal{P}(x) = J_{\Lambda}^{w} \mathcal{P}(\Lambda^{-1}(x)), \nonumber
\end{equation}
where $J_{\Lambda}$ is a Jacobian factor and the notation $\Lambda \cdot \mathcal{P}$ denotes the action of a superrotation $\Lambda$ on $\mathcal{P}$. The weight $w$ plays the role of the conformal dimension, but for the full diffeomorphism group. Another way of phrasing the dominant decomposition into hard and soft parts is to restrict to representations with a non-trivial global conformal weight. 

The stabilizers for each class of supermomenta may then be worked out in succession. The caveat is that the stabilizer is not strictly a gauge redundancy in the context of asymptotic symmetries. Large gauge transformations give rise to physically inequivalent configurations. Nevertheless, these configurations can be generated from each other by the action of the `stabilizer'. To that end, we make a distinction between a \textit{strict} stabilizer which stabilizes the supermomentum pointwise $\Lambda \cdot \mathcal{P} = \mathcal{P}$, and an \textit{asymptotic stabilizer} which stabilizes the \textit{class} of the supermomentum $\Lambda \cdot [\mathcal{P}] = [\mathcal{P}]$. Here, $[\mathcal{P}]$ can be one of three sets: constant, axisymmetric, or mixed smooth functions on the sphere.

All of this and more can be analyzed in detail, a task we delegate to Appendix~\ref{app: 1}. The main results of this work are succinctly summarized in Table~\ref{tab:supermomentum-stabilizers}.

\begin{table}[h!]
\centering
\renewcommand{\arraystretch}{1.15}
\begin{tabular}{@{}l@{\quad}l@{\quad}l@{}}
\toprule
\textbf{$\mathcal{P}(\theta, \phi)$} & \textbf{Strict} & \textbf{Asymptotic} \\
\midrule
Const.\ $\rho$ & $\text{SDiff}(S^2)$ & $\text{SDiff}(S^2) \rtimes \mathbb{R}^\times$ \\
Axi.\ $j(\theta)$ & $SO(2)_\phi \!\times\! G_\theta$ & $\text{Diff}(S^1)_\theta \rtimes SO(2)_\phi$ \\
Mixed $s(\theta,\phi)$ & Finite $G_{s} \subset SO(3)$ & Same as strict \\
\bottomrule
\end{tabular}
\caption{Classes of supermomenta $\mathcal{P}$ and their strict and asymptotic stabilizers under superrotations $G= \text{Diff}(S^2)$. Here, $G_{\theta} \in \{\text{trivial}, C_{n}\}$ , with $C_{n}$ being the cyclic group of order $n$, and $\mathbb{R}^{\times} = \mathbb{R} /\{0\}$. The vacuum is strictly and asymptotically stabilized by $\text{Diff}(S^2)$, as its constant representative is $\rho =0$. }
\label{tab:supermomentum-stabilizers}
\end{table}

The representations here exhaust all possibilities of classes of supermomenta acted on by superrotations according to our chosen decomposition. The fact that asymptotic stabilizers are infinite-dimensional suggests that the orbits are much smaller than those in the literature. Similarly to Wigner's classification of particles, one can attempt to distinguish the specific classes of excitations by labels arising from the orbit, like rest mass, and from the stabilizer, like spin. For the orbit with $\text{SDiff}(S^2)$ stabilizer, which is the analogue of the timelike particle with $SO(3)$ stabilizer, the labels from the (asymptotic) strict stabilizers give rise to (asymptotic) charges. These labels correspond to an infinite number of charges generalizing the $l$ and $m$ spin degrees of freedom of a massive particle (cf.~\cite{Penna:2018bzj}). Relatedly, a given orbit will generically contain an infinite number of representatives that are treated on the same footing relative to the asymptotic symmetry $B$. Note that supermomenta with a constant representative may be preserved pointwise, which is related to the fact that supertranslations do not affect the mass aspect. Under the action of a superrotation which is not in $\text{SDiff}(S^2)$, the mass aspect will change. This is the $\mathbb{R}^{\times}$ part of the asymptotic stabilizer.  In obtaining the axisymmetric orbit, we assume that $j(\theta)$ is axisymmetric and not constant, so one should think of those as having $\rho=0$. All of this is in direct analogy with the mass label of the Poincaré group, and the difference between stabilizers for timelike and null particles.

To summarize, the asymptotic excitations in Table~\ref{tab:supermomentum-stabilizers} resonate two well-known statements in the literature: (1) the infinite degeneracy of gravitational `vacua' or orbits and (2) the charges distinguishing between different representatives of a single orbit of an asymptotically flat spacetime under the action of the asymptotic symmetry $B$. Thus, they serve as building blocks for `particles' in a given spacetime, but also of an infinite number of asymptotically flat spacetimes. 
\section{Building up particles and spacetime} 
In this section, we contextualize the result of our work. This is done on several stages: (1) the concept of particles from Wigner, (2) the excitations of McCarthy, (3) the relevance of the asymptotic symmetry group $B$ in quantum gravity, (4) the difference between local conformal and `diffeomorphical' superrotations, and (5) the three-dimensional analogue of the asymptotic excitations. 

The first three points can be considered in tandem. One way of phrasing McCarthy's results on $B_{0}$ relative to those of Wigner on $P$ is the following: the (irreducible) excitations of $B_{0}$ become reducible excitations of $P$ upon restriction to a Poincaré subgroup. More physically, turning on supertranslations (which is group-theoretically encoded in extending the translation subgroup $\mathbb{R}^{4}$ to the supertranslation group $C^{\infty}(S^2)$) drastically changes the spectrum of the theory. Since turning on supertranslations amounts to changing the background from, say Minkowski to one of its related vacua~\cite{Compere_vacua_2016}, this means that a restriction to the Poincaré subgroup is essentially breaking the supertranslation asymptotic symmetry group and choosing a different background \textit{on top of which} excitations can propagate. Understanding the extent to which these excitations \textit{deviate from} Wigner's particles simply amounts to quantifying the misalignment of the stabilizers\footnote{In general, one can see this deviation at the level of a single orbit by distinguishing between an asymptotic and a strict stabilizer. }~$G_{\mathcal{P}}$ and $G_{p}$. Similarly, our work faithfully accounts for superrotations. The spectrum of a theory with superrotations turned on, which is encoded by extending from $L$ to $\text{Diff}(S^2)$, is drastically changed as well. To understand how these excitations deviate from either McCarthy's or Wigner's analyses again amounts to understanding the \textit{branching} of an asymptotic excitation under the restriction to a $B_{0}$ or a $P$ subgroup of $B$. The branching from $B_{0}$ to $P$ was worked out in Ref.~\cite{mccarthy2} and reinterpreted and expanded upon in Ref.~\cite{Bekaert:2025kjb} Physically, turning on superrotations additionally changes the original background spacetime~\cite{Compere_vacua_2016,Compere_schwarzchild_2016}, and this restriction amounts to choosing a particular one in its orbit. As $B$ is an asymptotic symmetry group, it is able to connect physically distinct spacetimes in quantum gravity. The concept of a particle, an excitation, and a spacetime become intermixed since one need not \textit{reduce to begin with.}

This already explains the reducibility, relative to $B$, of the excitations in the literature with orbits of the form
\begin{equation}
      O_{\mathcal{P}} \simeq  G_{p} / G_{\mathcal{P}} \times  L / G_{p}, \nonumber
\end{equation}
which are the explicit subject of Refs.~\cite{mccarthy0,mccarthy1,mccarthy2,mccarthy3,McCarthy_75, McCarthy_76_IV,McCarthy:1974aw, McCarthy_78,McCarthy_78errata,bekaert2024bmsparticles, Bekaert:2025kjb}, and the implicit subject of many investigations related to asymptotically flat spacetimes. The reason for this reducibility is the infinite-dimensional space arising from the superrotations, namely $\text{Diff}(S^2)/ L$, which is independent of the structure of $\mathcal{P}$. To exemplify this in gravity, consider starting with global Minkowski $\eta$. Turning on supertranslations generates an infinite degeneracy, labeled by a function on the sphere $C$, which generates the orbit $O_{\eta}^{0}$ relative to $B_{0}$. Turning on superrotations has a similar effect, and generates the orbit $O_{\eta}$ relative to $B$, which is labeled by $C$ and a Weyl factor $\Omega$. These are the avatars of the supertranslation and superrotation fields of Ref.~\cite{Compere_vacua_2016}. In our context, more fine-grained information emerges. 

Consider an asymptotic excitation with constant density $\rho$. The orbit is preliminarily labeled by $\text{Diff}(S^2)/ \text{SDiff}(S^2) \simeq \mathcal{W}$. The Lie algebra of the stabilizer is given by divergenceless vector fields $Y^{A}$ on the sphere. Alternatively, these integrate up to traceless symmetric tensors $C_{AB}$, if the weight $w$ is chosen such that the supermomentum is spin-weighted~\cite{Barnich:2021dta}. Such objects decompose into 
\begin{equation}
    C_{AB} = C^{0}_{AB} + \hat{C}_{AB}, \nonumber
\end{equation}
where the first is given by a single scalar $C(\theta, \phi)$ on the sphere
\begin{equation}
     C^{0}_{AB}: =  \left[ D_{A} D_{B} -  \frac{1}{2} \gamma_{AB} D^2 \right]C. \nonumber
\end{equation}
To understand the second, we must look at additional labels of the asymptotic excitation. The label $C(\theta, \phi)$ naturally arose from the orbit, the remaining ones will emerge from the spin space $\mathfrak{h}$. In analogy to $l$ and $s$ spin labels of a timelike particle, we turn to the Lie algebra comprised now of divergenceless vector fields $N^{A}$. Such objects decompose as
\begin{equation}
    N^{A} = \epsilon^{AB} D_{B} \Phi, \nonumber
\end{equation}
where $\Phi$ is some function on the sphere, referred to as the vorticity potential in hydrodynamics since it obeys $\gamma = D^{2} \Phi$ where $\gamma$ is the vorticity function. This means that the second factor describing the asymptotic excitation, $\hat{C}_{AB}$, may further be decomposed as
\begin{equation}
  \hat{C}_{AB} = \epsilon^{C}_{(A}D_{B)}D_{C}\Phi.  \nonumber 
\end{equation}
What we have learned is that the simplest asymptotic excitation is able to be parametrized by precisely the supertranslation and superrotation fields appearing in the context of orbits of gravitational spacetimes. Our results assume one works in a Hilbert space, so the unitary version of the asymptotic excitations will be the quantum description of this classical gravitational object. Another way of delineating $C^{0}$ from $\hat{C}$ is by realizing that they are simply the \textit{electric} and \textit{magnetic} components of $C_{AB}$, in the sense of partity. This explains the existence of different types of conserved charges in the literature~\cite{iyer1994some, wald2000general, barnich2002covariant, barnich2010finite,barnich2011bms, barnich2012note, flanagan2017conserved, Compere_phase_2018, grant2022wald, odak2023wald}.

On the last two points, one can instead reconsider our analysis by taking $G$ to be the local conformal group given by \textit{local conformal superrotations}. This would replace the degeneracy label with $\text{Conf}(S^2)/ L$ instead. The spectrum of the theory is modified already, since this assumption would force the stabilizers to live in the part of $\text{Diff}(S^2)$ which rescales the Weyl factor. These include the $\text{SDiff}(S^2)$ transformations which are also conformal, which can be identified with $SO(3)$\footnote{Any local conformal transformation satisfies $f^{*} \gamma = e^{2\sigma} \gamma$, where $\gamma$ is the round metric on the sphere. For this to preserve the constant supermomentum, the Weyl factor $\sigma$ must be locally zero. This forces $f \in SO(3)$; that is, $f$ must be an isometry. A similar statement holds in $\{z, \bar{z}\}$ coordinates.}. They may also mix with the Weyl subgroup $\mathcal{W}$. This heightens the tension between conformal or general diffeomorphisms as candidates for the definition of superrotations. It would also align with McCarthy's classification of the stabilizers of $B_{0}$, which were necessarily compact subgroups of $SO(3)$. 

Finally, we can apply our analysis involving the asymptotic excitation stabilized by $\text{SDiff}(S^2)$, in three-dimensional gravity. The stabilizer turns out to be $\text{SDiff}(S^1) \simeq SO(2)$. This is precisely the massive `particle' obtained in Refs.~\cite{barnich2014notes, barnich2015notes, Oblak_thesis_2016}. Since three-dimensional gravity has no propagating degrees of freedom, there is no distinction between asymptotic excitations with infinite-dimensional stabilizers and excitations or `particles' with finite ones instead. 
\section{Prospects} 
The story we have told above is kinematical but can be promoted to a dynamical description by obtaining the action and charges of the representations~\cite{Souriau1970,Kostant1970,Kirillov1976,Kirillov2004,Barnich_actions_2018, Barnich:2022bni}. For the  asymptotic excitations with a $\text{SDiff}(S^2)$ stabilizer, we expect a fluid-like theory. For those with the axisymmetry, we expect a Schwarzian-like theory in line with the three dimensional result~\cite{Cotler:2024cia}. In principle, one can attempt to prepare these excitations in the lab, by engineering the specific Hamiltonian obtained from this analysis. The orbit associated to the former excitations is the space of Weyl factors on the sphere, and thus may naturally be related to Liouville theory for the celestial metric and the superrotation defects~\cite{Compere_vacua_2016}. We also showed that, with the right weight, these asymptotic excitations can be thought of, classically, as diffeomorphic but physically distinguishable geometries by way of an infinite number of charges.  It would be of interest to relate these excitations to soft theorems and the memory effect through the infrared triangle~\cite{Strominger:2017zoo}. The decomposition used here already reveals the electric and magnetic components of an asymptotically flat spacetime, either of which may be used to construct conserved charges.

\begin{figure}[!ht]
    \centering
    \includegraphics[width=0.7\linewidth]{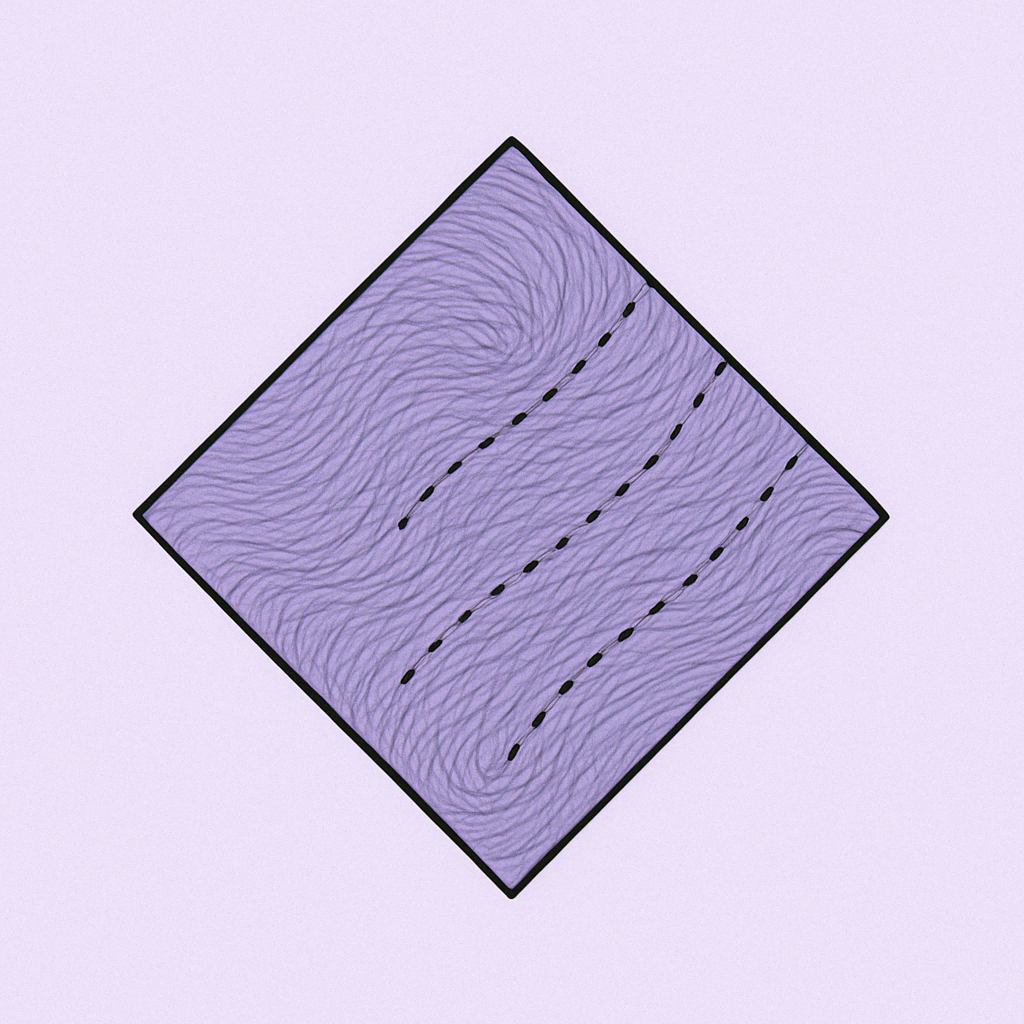}
    \caption{The asymptotic excitations are visually displayed in a spacetime with only the asymptotic structure at null infinity fixed. The Penrose diagram becomes more certain as we approach the boundary, but the dynamics of asymptotic excitations can distort the geometry away from Minkowski. The worldlines represent reducible propagating excitations. If the asymptotic excitations localize on the boundary, the wavy interior of the diagram should be interpreted as bulk imprints of boundary dynamics. }
    \label{wavy penrose}
\end{figure}

In the context of quantum gravity, one can interpret the asymptotic excitations as prescribing the dynamics of spacetime itself. Adding more structure to the supermomentum constrains the stabilizer to become finite-dimensional, and eventually trivial. We pointed out the distinction between excitations that make up spacetime, and those that propagate within a given background after reduction to special subgroups of $B$. Even after this violence, the particles remember the broken symmetry in their configuration space structure. Thus, applying the inducing construction to an asymptotic spacetime symmetry group breaks the distinction between propagating matter and dynamical gravity, cf.~Figure~\ref{wavy penrose}. 

In an attempt to further label these new excitations, one may extend the symmetry even more to obtain additional charges. It is well-known that the only non-trivial extensions of the Poincaré group with internal continuous symmetries must be direct products~\cite{coleman1967, haag1975}. There may be ways to extend the group $B$ non-trivially to $\hat{B}$, that bypass the no-go theorems, though see Ref.~\cite{Fuentealba:2023hzq} for an analogous no-go statement. This extension would also be important for the study of different theories, that are not asymptotically flat quantum gravity, but still share the same symmetry group with potentially different \textit{internal} charges.

Restricting gravitational dynamics to a subregion should also break the symmetry group to the corner symmetry subgroup, and in this context, one expects that the `corner' excitations will be of the spacetime region itself instead of those propagating within it. It is natural to wonder if these asymptotic excitations, or their corner analogues, are codimension-one, or boundary, modes in four dimensions. 

If a celestial holography exists with the same symmetry group $B$~\cite{Strominger:2017zoo, Pasterski:2021raf}, a natural guess would be that further analysis of these representations should inform us about the basic structure of the putative dual theory. For example, this could be the Carrollian conformal field theories placed on null boundaries~\cite{Duval:2014uva,Donnay:2019jiz}. However, it may be the case that these setups force us to consider conformal stabilizers and neglect the full superrotation symmetry. It would also be of independent interest to construct quantum reference frames for fixed gravitational asymptotics so that our representations can play the role of observers in asymptotically flat space~\cite{Kabel:2023jve}. 

Relatedly, we highlight a tension between two potential definitions of a superrotation in four dimensions. In three dimensions, one can compute the one-loop partition function for pure gravity independently, and match it to the vacuum character of the BMS group in three dimensions~\cite{Barnich:2015mui, Oblak:2015sea}. Diffeomorphisms and local conformal transformations are the same entity in three dimensions. In four dimensions, one could answer the question by following a similar strategy. We hope to report on the outcome in future work.

It is also of interest to understand our classification in the context of topologies generally described by Riemann surfaces~\cite{Barnich:2021dta}. This would be relevant for topology change in quantum gravity.  We have already seen that taking superrotations to be diffeomorphisms reveals a richer theory, despite the necessary foray into non-conformal ground. The global conformal transformations for higher genus $g$ beyond the torus trivialize, and thus there is no recourse to an analogous `Lorentz subgroup' anymore. Beyond $g>1$, it seems that only asymptotic excitations will remain.

Finally, we come to thermality and black holes. The question of unitary evolution for evaporation may be revisited now that the spectrum of the theory is known. Take Schwarzchild as an example.  Other spacetimes in Schwarzchild's orbit under superrotations, for example, only differ up to refined charges~\cite{Compere_schwarzchild_2016, Compere:2016gwf}. These charges are sensitive to the interior, suggesting that a subclass of representations found here can be used to label potential microstates of the black hole.  If in the process of evaporation, a black hole transitions to another spacetime in the orbit, then the excitations in the dominant semiclassical geometry could carry information allowing the reconstruction of the star which collapsed to form it~\cite{Hawking:2016sgy}. The inclusion of the asymptotic orbit of a saddle in the gravitational path integral may influence the thermodynamic stability of asymptotically flat spacetimes. In future work, we consider these questions in detail.

In anti-de Sitter space, a similar phenomenon occurs for black holes where a transition between competing gravitational saddles helps recover some aspects of unitarity. The non-trivial saddles, which are typically called replica wormholes, may be reinterpreted in terms of the original saddle with \textit{defects}, recovering the area term in the entropy\footnote{It is important to understand the effect these representations have on entropy and how they fuse together. One expects some divergences from the large stabilizers, which may be cured by appropriate truncation to certain subgroups~\cite{Penna:2018bzj}} of large black holes~\cite{Almheiri:2020cfm}. This evokes the story told in Ref.~\cite{Compere_vacua_2016}, where the other elements in the orbit of flat space under superrotations have spinning defects, though see Ref.~\cite{Adjei:2019tuj} for an alternative perspective.

All these interactions reveal that the representation theory of asymptotic spacetime symmetries provide an organizing principle to understanding quantum gravity with fixed asymptotics, and beyond. In fact, the symmetry group considered here can be obtained by the flat limit of a more general group that depends on the sign of the cosmological constant~\cite{Compere:2019bua}. In de Sitter, the supertranslations no longer form an abelian subgroup and so a new idea may be needed. It may even be the case that $B$ is only a subgroup of a larger symmetry we have not yet uncovered, thereby making our asymptotic excitations reducible upon restriction. We believe that the perspective of inducing excitations for asymptotic symmetries will be crucial in understanding quantum gravity.

\section*{Acknowledgements} We are grateful to many people, including: Ahmed Almheiri, Gerald Goldin, Gaston Giribet, Greta Goldberg, Marc S.~Klinger, Monica Pate, and Yifan Wang. 

\providecommand{\noopsort}[1]{}\providecommand{\singleletter}[1]{#1}%

\appendix
\onecolumngrid
\section{Asymptotic stabilizers} \label{app: 1}
In this section, we work out the stabilizers of the three classes of supermomenta tailored to the decomposition of the space of smooth functions on the sphere. We will not make an assumption on the weight $w$ of the supermomentum until we need to. For convenience, the decomposition is 
\begin{equation}
    C^{\infty}(S^2) = \mathbb{R} \oplus C^\infty_{\text{axi}}(S^2) \oplus C^\infty_{\text{mixed}}(S^2),
\end{equation}
while the action on the supermomentum $\mathcal{P}$ is
\begin{equation}
    (\Lambda \cdot \mathcal{P})(x) := J^w_\Lambda(x) \mathcal{P}(\Lambda^{-1}(x)),
\end{equation}
where $J_\Lambda(x)$ is the Jacobian determinant of the diffeomorphism $\Lambda$ at the point $x \in S^2$, and $w$ is the conformal weight of $\mathcal{P}$. To determine the stabilizers, we simply solve the differential equation associated to the stabilization condition
\begin{equation}
    (\Lambda\mathcal{P})(x) = \mathcal{P}(x).
\end{equation}
In fact, there are two natural stabilization conditions to consider. The first is that of strict stabilization, as in, the above equation must hold pointwise. The second is that of asymptotic stabilization, which entails only fixing the \textit{class} of the supermomentum. We will perform both simultaneously, but we find it more natural to fix the class as the $B$ symmetry is an asymptotic symmetry unlike the Poincaré group. This means that it in principle can change the charges of a given configuration, and still entertain physically inequivalent yet diffeomorphic configurations within the same orbit.

The procedure is to obtain a group formed by specific $\Lambda$'s under the assumption that $\mathcal{P}$ belongs to any of the three classes of the decomposition. However, since we have more space in the appendix, we will derive the classes from the structure of the differential equation itself.

Namely, we will striate the classes of solutions by the degree of simplicity the equation takes, or equivalently, the degree of symmetry it possesses upon this restriction. Broadly, one could have an infinite-dimensional symmetry or a finite one. The larger the symmetry, the easier the equation is to solve. To that end, we split our analysis into two general classes admitting
\begin{enumerate}
    \item \textbf{Infinite-dimensional symmetries:} 
    \begin{itemize}
        \item \textit{Constant class} The simplest equation to solve would be if the orbit of $\mathcal{P}$ can be chosen to have a constant representative, that is $O_{\mathcal{P}} \ni \rho \in \mathbb{R}$. In that case, the \textit{strict} stabilization condition becomes
    \begin{equation}
        \Lambda\cdot \mathcal{P} = J_{\Lambda}^{w}(x) \mathcal{P} \iff J_{\Lambda}^{w}(x) \rho = \rho \iff J_{\Lambda}^{w}(x) = 1 \iff G_{\rho} = \text{SDiff}(S^2).
    \end{equation}
    As long as $w \neq 0$, this would only hold for superrotations that belong to $\text{SDiff}(S^2)$, which satisfy $J_{\Lambda}(x)=1$ for all $x \in S^2$. If $w =0$ for $\mathcal{P}(x) = \rho$, then this would hold for any diffeomorphism in $\text{Diff}(S^2)$. Recall that $w$ is the weight of the supermomentum, which is typically a density (not a scalar) even in three-dimensions~\cite{Oblak_thesis_2016}. The stabilization condition in this case for a physical supermomentum that can be put to rest enjoys the highest degree of symmetry, as reflected by the infinite-dimensionality of both $\text{SDiff}(S^2)$ and $\text{Diff}(S^2)$. The former is the \textit{asymptotic stabilizer} for $\rho \neq 0$, while the full group of superrotations $G= \text{Diff}(S^2)$ stabilizes the special case of $\rho =0$ even when $w \neq 0$. 

    The asymptotic stabilization condition would differ slightly from the above analysis. It is no longer a question of fixing the value $\rho$ itself for a given supermomentum, but of remaining within the class of supermomenta admitting \textit{any} constant representative value. This means that the asymptotic stabilizer would be an $\mathbb{R}^{\times}$ extension\footnote{The notation means $\mathbb{R}^{\times} = \mathbb{R} - \{0\}$; non-zero real numbers.} of $\text{SDiff}(S^2)$. In other words, the asymptotic stabilizer for the constant class is 
    \begin{equation}
        G_{\rm cst}^{\rm asymptotic} = \text{SDiff}(S^2) \rtimes \mathbb{R}^{\times} := \{ \Lambda \in \text{Diff}(S^2); J_{\Lambda} = \text{cst}\}. 
    \end{equation}

    The stabilization equation simply isolated a nice condition on $J_{\Lambda}$ in this case, because of the large asymptotic stabilizer. It is not yet clear if there admits an infinite-dimensional stabilizer which is not $\text{SDiff}(S^2)$, that leads to a less obvious condition on $\Lambda$. One way of phrasing the above derivation is by removing the functional dependence of $\mathcal{P}(x)$ on the entire sphere, so that locally it is a constant and thus independent of the two coordinates $(\theta, \phi)$. 

    \item Motivated by this, one can then try to go halfway by simply considering the stabilization condition when the orbit admits an \textit{axisymmetric} representative $j(\theta)$. In that case, the strict stabilization condition reads
    \begin{equation}
        J^{w}_{\Lambda}(\theta, \phi) j\left[\Lambda^{-1}(\theta,0)\right] = j(\theta).
    \end{equation}
    We have made the independence on $\phi$ explicit in $j$, since it is a function on the two-sphere enjoying an axisymmetry, and not simply a function on the circle. The fact that $\Lambda$ also acts on the argument will force $j$ to no longer be axisymmetric, which means that for the above condition to hold, it cannot reparametrize $\phi$ arbitrarily. Indeed, it can only rigidly rotate it, so $\Lambda$ will at most effect a transformation in $SO(2)_{\phi}$. This also eliminates the $\phi$ dependence of the Jacobian factor. To preserve the \textit{specific} axisymmetric supermomentum, $j(\theta)$, point-wise, we see that the strict stabilizer is at most
    \begin{equation}
        G_{j} = SO(2)_{\phi} \times G_{\theta},
    \end{equation}
    where $G_{\theta} \leqslant SO(2)_{\theta}$ is a subgroup of rigid rotations of $\theta$, depending on the symmetry of the given $j(\theta)$. This could be trivial if $j$ is a generic function of $\theta$, a discrete rotational symmetry $C_{n}$ if $j(\theta)$ satisfies
    \begin{equation}
        j(\theta) = j(\theta + \frac{2\pi}{n}),
    \end{equation}
    and thus can be expanded in terms of $\cos n k \theta$, and in principle could be $SO(2)_{\theta}$ itself. The issue with the last group is that this would force $j(\theta)$ to be independent of $\theta$, and thus the supermomentum would have admitted a constant representative to begin with. Thus, this possibility is excluded. 

    Of course, we are interested in infinite-dimensional symmetries still. Despite the strict stabilizers being finite-dimensional for the axisymmetric class, the asymptotic stabilizers will not be. Let us see what occurs when we wish to remain in the same axisymmetric class, but allow different $j(\theta)$'s in a single orbit. The fact that $\Lambda \in \text{Diff}(S^2)$ also acts on the argument will generically make the orbit of $j(\theta)$ depend on $\phi$. The same argument above isolates the $\phi$ part of the symmetry to be $SO(2)_{\phi}$. Since we are treating momenta as densities ($w \neq 0$), this suggests that one need only preserve this condition under integration. Thus, this condition holds if and only if $\Lambda \in \text{Diff}(S^1)_{\theta} \rtimes SO(2)_{\phi}$, as a reparametrization of $\theta$ will be counteracted by the rescaling of the area element on the two-sphere. The asymptotic stabilizer is then
    \begin{equation}
        G_{\rm axi}^{\rm asymptotic} = \text{Diff}(S^1)_{\theta} \rtimes SO(2)_{\phi}, 
    \end{equation}
    which \text{is} an infinite-dimensional symmetry. 
    \end{itemize}

    \item \textbf{Finite-dimensional symmetries:} At this point, we have exhausted the infinite-dimensional symmetries since there is no more recourse to removing angular dependence on the sphere. Indeed, for a generic \textit{mixed} supermomentum $s(\theta, \phi)$, that depends (locally) on the two angles, the stabilizer may very well be trivial. Before jumping to that conclusion, we must analyze finite-dimensional stabilizers in $\text{Diff}(S^2)$ of which the trivial subgroup is one example. The more generic $\mathcal{P}$ is, the more constrained the symmetry is;  the larger the stabilizer, the more constrained the supermomentum is. Fortunately, we may borrow elements of McCarthy's original analysis~\cite{mccarthy2} while introducing some novel mathematical reasoning.

    To understand this class, we may search for subgroups of $\text{Diff}(S^2)$ that are not infinite-dimensional in the same way that $\text{Diff}(S^1)$ is. The biggest such symmetry would be a Lorentz subgroup $L= SO(3,1)$. Now, any smooth function on the sphere strictly stabilized by $L$ would have to be (locally) constant. This possibility is excluded. Thus, we need to search for finite-dimensional subgroups of $L$. The maximal one would be $SO(3)$. Thus, we arrive at a similar statement to McCarthy's for the $B_{0}$ symmetry, but for the full asymptotic symmetry group $B$ in a nice synthesis: any supermomentum $\mathcal{P}$ with a \textit{mixed} representative $s(\theta, \phi)$ can only be stabilized by a compact subgroup of $SO(3)$
    \begin{equation}
        G_{s} \leqslant SO(3), \quad \text{$G_{s}$ is compact}.
    \end{equation}

    Now, if $G_{s} = SO(3)$, the mixed function $s$ must be locally constant. This is excluded. In the same, $G_{s}= SO(2)$ is excluded because it would then be an axisymmetric function. The only possibilities are now \textit{discrete subgroups of $SO(3)$. } These are classified and for details we ask the reader to consult~\cite{mccarthy2}. In sum, these could be $C_{n}, D_{n}, T, O,$ or trivial. Equivalently and respectively, these are the cyclic, dihedral, tetrahedral, octahedral, icosahedral groups which are the symmetries of $n$-polygons.

    To tie up the final loose end, one may ask about the asymptotic stabilizers of the mixed class. Recall that the constant and axisymmetric classes enjoyed a large degree of continuous degeneracy: scaling to any constant function while remaining in the same orbit, and reparametrizing the function $j(\theta)$ itself. The mixed class does not enjoy this ambiguity, because the constraints on its representative's functional dependence are too strong to be continuously deformed. All this to say: the asymptotic and strict stabilizers of the mixed class are one and the same. 
\end{enumerate}

\end{document}